\documentclass[apjl]{emulateapj}
\usepackage{epsfig}
\begin{document}

\title{Constraining spiral structure parameters through Galactic pencil-beam and 
large-scale radial velocity surveys}

\author{I.~Minchev\altaffilmark{1}
and A.~C.~Quillen\altaffilmark{1}}
\altaffiltext{1}{Department of Physics and Astronomy, University of Rochester, 
Rochester, NY 14627; iminchev@pas.rochester.edu, aquillen@pas.rochester.edu}

\begin{abstract}
We investigate the effect of spiral structure on the Galactic disk
as viewed by pencil beams centered on the Sun, relevant to upcoming
surveys such as ARGOS, SEGUE, and GAIA. We create synthetic Galactic maps 
which we call Pencil Beam Maps (PBMs) of the following observables: 
line-of-sight velocities, the corresponding velocity dispersion, 
and the stellar number density that are functions of distance from the observer.
We show that such maps can be used to infer spiral structure parameters, such as
pattern speed, solar phase angle, and number of arms. The mean line-of-sight 
velocity and velocity dispersion are affected by up to $\sim35$ km/s
which is well within the detectable limit for forthcoming radial velocity surveys. 
One can measure the pattern speed by searching for imprints of resonances.
In the case of a two-armed spiral structure it can be inferred 
from the radius of a high velocity dispersion ring situated at the 2:1 ILR. 
This information, however, must be combined with information related to 
the velocities and stellar number density in order to distinguish from a 
four-armed structure.
If the pattern speed is such that the 2:1 ILR is hidden inside the Galactic bulge
the 2:1 OLR will be present in the outer Galaxy and thus can equivalently be 
used to estimate the pattern speed. Once the pattern speed is known the solar angle
can be estimated from the line-of-sight velocities and the number density PBMs.
Forthcoming radial velocity surveys are likely to provide powerful constraints of
the structure of the Milky Way disk.

\end{abstract}


\section{Introduction}
It has been well established by now that the Milky way is not axisymmetric
with both a central bar and spiral structure perturbing its disk. 
Due to our location in the Galactic plane both spiral and bar structure is
impossible to observe directly. Galactic bar parameters such as orientation 
and pattern speed have been inferred indirectly from both asymmetries around the 
Galactic center (e.g., \citealt{blitz91,weinberg92})
and its effect on the local velocity distribution of old stars, i.e., the 
Hercules stream \citep{dehnen99,dehnen00,fux01,mq07b}.

Spiral structure parameters, however, are much more uncertain. Current spiral 
density wave models \citep{fux01,lepine01,desimone04,qm05} strongly disagree 
on the strength of the spiral structure, the number of arms, and the 
pattern speed. These models differ in their predictions of the 
induced velocity streaming at different angular positions in the Galaxy. 
For example, a four-armed density wave with velocity perturbations of 
$\sim20$ km/s will exhibit rapidly varying radial and tangential velocity 
components with azimuth across distances of a few kpc, and we could expect 
to detect $\sim20-50$ km/s variations in the mean line-of-sight stellar 
velocity as a function of the distance from the Sun. However, the strength of 
the spiral arm perturbation remains controversial. Based on the OGLE number 
counts, \cite{paczynski94} estimated that the Sagittarius-Carina arm has a factor 
of two increase in density compared to the underlying disk. This model is 
inconsistent with COBE studies which find a much smaller contrast ($\sim15\%$) 
and show that the Perseus and Scu-Cru arms are more dominant \citep{drimmel01}.

HI, CO, Cepheid, and far-infrared 
observations suggest that the Galactic disk contains a four-armed tightly 
wound structure. On the other hand, \cite{drimmel01} have shown that the
near-infrared observations are consistent with a dominant two-armed structure.
\citet{lepine01} suggest that locally the Milky Way can be modeled by the 
superposition of a two- and four-armed structure moving at the same pattern speed. 
By studying the nearby
spiral arms, \cite{naoz07} find that the Sagittarius-Carina arm is a superposition of
two features, moving at different pattern speeds. The effect of a two- and four-armed
structure, moving at different angular velocities, on the velocity dispersion
of a galactic disk has been explored numerically by \cite{mq06}.

Estimates for the pattern speed of the Milky Way spiral structure, or equivalently, 
the Sun's position with respect to resonances associated with spiral structure, 
span a large range of values. Reviewing previous work, \citet{shaviv03} finds 
a clustering of estimates for the pattern speed of local spiral structure near
$\Omega_s \sim  20 {\rm  km s^{-1} kpc}^{-1}$, though other studies suggest 
$\Omega_s \sim 13 {\rm  km s^{-1} kpc}^{-1}$. The model by \citet{lepine01} 
places the Sun near the corotation resonance 
$\Omega_s \sim  28 {\rm  km s^{-1} kpc}^{-1}$), and was fit to Cepheid kinematics.
The recent gas dynamical studies \citep{martos04, bissantz03}
match the properties of the gas in nearby arms with a spiral pattern speed of 
$\sim 20 {\rm km s^{-1} kpc}^{-1}$. \citet{martos04} propose that a two-armed 
stellar structure consistent with the stellar distribution inferred from COBE
could cause four-arms in the gas distribution near the Sun. The gas dynamical 
model proposed by \citet{bissantz03} with a similar spiral pattern speed 
matches HI and CO kinematics. The pattern speed of a spiral density wave can 
be tightly constrained from the location of its resonances. For example, 
\cite{qm05} associated stellar streams in the solar neighborhood with the 4:1 
ILR resonance of a two-armed pattern and were then able to tightly constrain 
the pattern speed of the driving spiral density wave to within 5\%.
Independent constraints on the pattern speed come from recent surveys of nearby 
open clusters (e.g. \citealt{dias05}) where the older clusters are found to have 
drifted further from their original density wave location. These authors concluded
that the Sun is located near the CR. A solar circle near the CR is also favored 
by \cite{lepine01} and \cite{naoz07}.  

In this paper we investigate how spiral structure parameters can be inferred
from velocity and density maps resulting from pencil-beam and large-scale 
surveys of the Galaxy. 
At present the influence of spiral arms on the observed kinematic properties of the 
Galactic disk is very poorly understood. With the advent of 
future Galactic all-sky (GAIA, SEGUE) and pencil-beam (ARGOS, BRAVA) radial 
velocity surveys, large amounts of kinematic data will be collected. 
The types of dynamical constraints made possible with these new data sets is not 
currently known. We address that issue here with synthetic models for the purpose of
exploring how spiral structure might be constraint from these data.

\section{The simulations}

\begin{figure*}
\epsscale{1.2}
\plotone{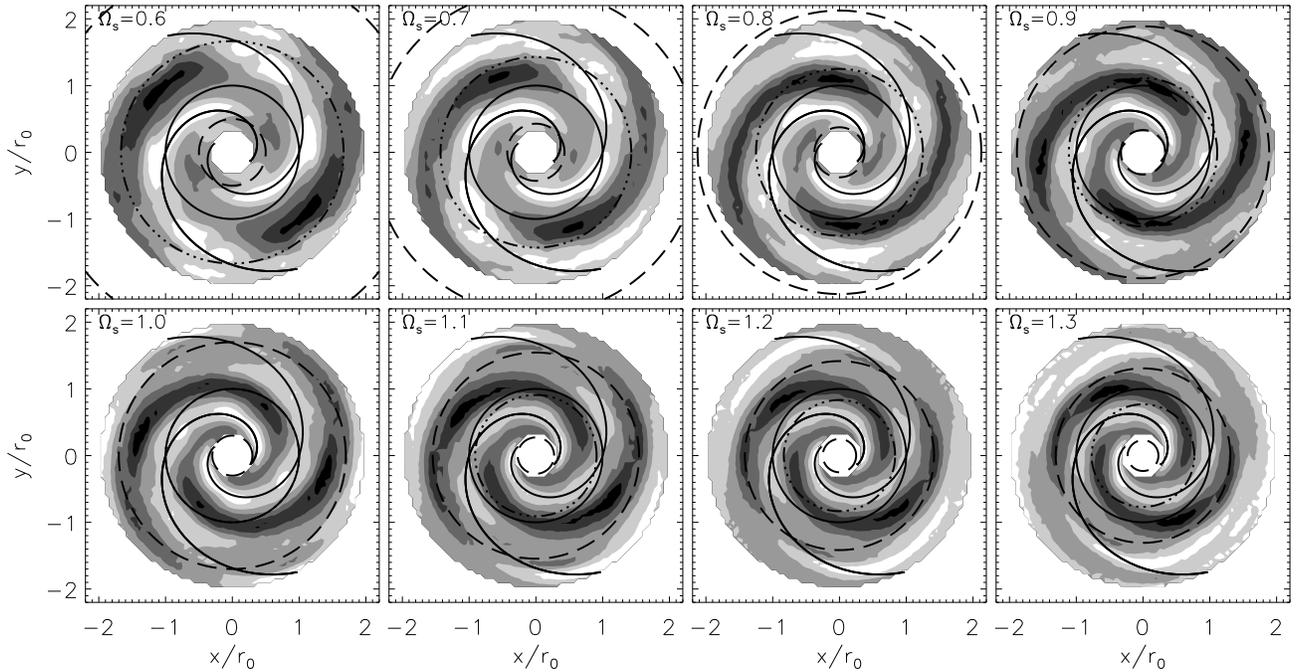}
\figcaption{
Each panel shows stellar number density contour plots of a simulation with a 
particular spiral pattern speed, $\Omega_s$, in units of the local circular velocity, 
$\Omega_0$. The axisymmetric disk is subtracted to emphasize the perturbation. 
The quantity plotted is $(\Sigma-\Sigma_{axi})/\Sigma_{axi}$, where $\Sigma$ and
$\Sigma_{axi}$ are the perturbed and axisymmetric stellar number densities.
Contours are spaced linearly with white indicating highest values.
Dashed circles show 2:1 LRs, the dash-dotted line indicates the CR, and the solid 
circle represents the solar neighborhood at $r/r_0=1$. The inner $0.3r_0$ is not 
plotted since we do not simulate the central bar. We consider pattern speeds between
the 4:1 ILR and the 4:1 OLR. 
\label{fig:den_m2}
}
\end{figure*}

We perform 2D test-particle simulations of an initially axisymmetric exponential 
galactic disk. In order to reproduce the observed kinematics of the Galactic disk, 
we use disk parameters consistent with observations (table \ref{table:par}). 
The reader is referred to \cite{mq07a} for a more detailed description of our
simulation set up.
In all of our simulations we start with an initially warm disk, i.e., 
the radial velocity dispersion at $r_0$ is $\sigma_u=0.20v_0$ where $v_0$ is the 
velocity of the local standard of rest.
The background axisymmetric potential due to the disk and halo has the form 
$\Phi_0(r)=v_0^2\log(r)$, corresponding to a flat rotation curve.

We treat the spiral pattern as a small perturbation to the axisymmetric
model of the galaxy
by viewing it as a quasi-steady density wave in accordance with the Lin-Shu
hypothesis \citep{lin69}. The spiral wave gravitational potential
perturbation is expanded in Fourier components as
\begin{equation}
\label{eq:sp}
\Phi_1(r, \phi, t)=\sum_m \epsilon_m
\exp \left[i (\alpha \ln{r} - m(\phi-\Omega_s t))\right].
\end{equation}
The parameter $\alpha$ is related to the pitch angle of the spiral wave, 
$p$, as $\alpha = m \cot(p)$, negative for trailing spirals with rotation
counterclockwise, and $(r,\phi)$ are plane polar coordinates. The pattern
speed is given by $\Omega_s$ and the spiral strength by $\epsilon_m$.
For a two-armed structure the $m=2$ term dominates.
Upon taking the real part of equation \ref{eq:sp}
the perturbation due to the two-armed spiral density wave becomes

\begin{deluxetable}{lcc}
\tablewidth{3in}
\tablecaption{Simulation parameters used\label{table:par}}
\tablehead{
\colhead{Parameter}                  &
\colhead{Symbol}                    &
\colhead{Value}           
}
\startdata
Solar neighborhood radius   & $r_0$  &   1                            \\
Circular velocity at $r_0$  & $v_0$  &   1                            \\
Radial velocity dispersion & $\sigma_{\rm u}(r_0)$ & $0.20v_0$ \\  
$\sigma_{\rm u}$ scale length & $r_{\sigma}$ & $0.9r_0$                     \\
Disk scale length  &  $r_\rho$       &    $0.37r_0$                    \\
Spiral strength    &  $\epsilon_{\rm s}$   &     $-0.015$              \\
Pitch angle        &   $p$         &     $18^\circ$      
\enddata
\end{deluxetable}

\begin{equation}
\Phi_1(r,\phi,t) =
\epsilon_s \cos{(\alpha \ln{r}-2(\phi-\Omega_s t))}.
\end{equation}

Integrations are performed forward in time. The perturbation is grown from
zero to its maximum strength in four rotation periods at $r_0$.
In order to improve statistics, positions and velocities are time averaged 
for 10 spiral periods. We distribute particles (stars) between in inner and
outer galactic radii $(r_{in}, r_{out})=(0.3r_0,2.0r_0)$. New particles are
added until the final number of outputs is $2.5\times10^6$. In addition,
the two-fold symmetry of our model galaxy is used to double this number.

We present our results by changing the spiral pattern speed, $\Omega_s$,
and keeping the solar radius fixed at $r_0=1$.
For a two-armed spiral pattern the primary resonances are the 2:1 
inner and outer Lindblad resonances (ILR and OLR). Those are achieved when  
$\Omega_s/\Omega_0=1+\kappa/2\approx0.3, 1.7$, respectively, 
where $\kappa$ is the epicyclic frequency. Similarly, the second order 
resonances are the 4:1 Lindblad resonances (LRs) at 
$\Omega_s/\Omega_0=1+\kappa/2\approx0.65, 1.35$.
We examine a region of parameter space for a range of pattern 
speeds placing the SN between the 4:1 LRs.

\section{Variation of Galaxy morphology with pattern speed}

Interpretation of line-of-sight velocities with Galactic longitude and distance
from the Sun is not straightforward. To help out we first discuss morphology as
seen by an outside viewer. 

In figure \ref{fig:den_m2} we present stellar number density contour plots for 
simulations of galactic disks with different pattern speeds. The background 
axisymmetric disk is subtracted to emphasize the spiral structure. 
The quantity plotted is $(\Sigma-\Sigma_{axi})/\Sigma_{axi}$, where $\Sigma$ and
$\Sigma_{axi}$ are the perturbed and axisymmetric stellar number densities.
Concentric circles
represent the 2:1 LRs (dashed), the solar radius (solid), and the CR (dash-dotted).
Darker colors correspond to lower density. The inner $0.3r_0$ disk is not plotted
since we do not model the Galactic center. Each panel represents a simulation with
a distinct pattern speed, $\Omega_s$, and all other parameters kept the same (see 
table \ref{table:par}). Pattern speeds considered range approximately between
the 4:1 LRs, $\Omega_s=[0.6,1.3]\Omega_0$ in units of 0.1.  
The minima of the two-armed spiral potential are graphed in each panel as solid 
curves. Note the crowding of resonances as the pattern speed increases.

\begin{figure*}
\epsscale{1.0}
\plotone{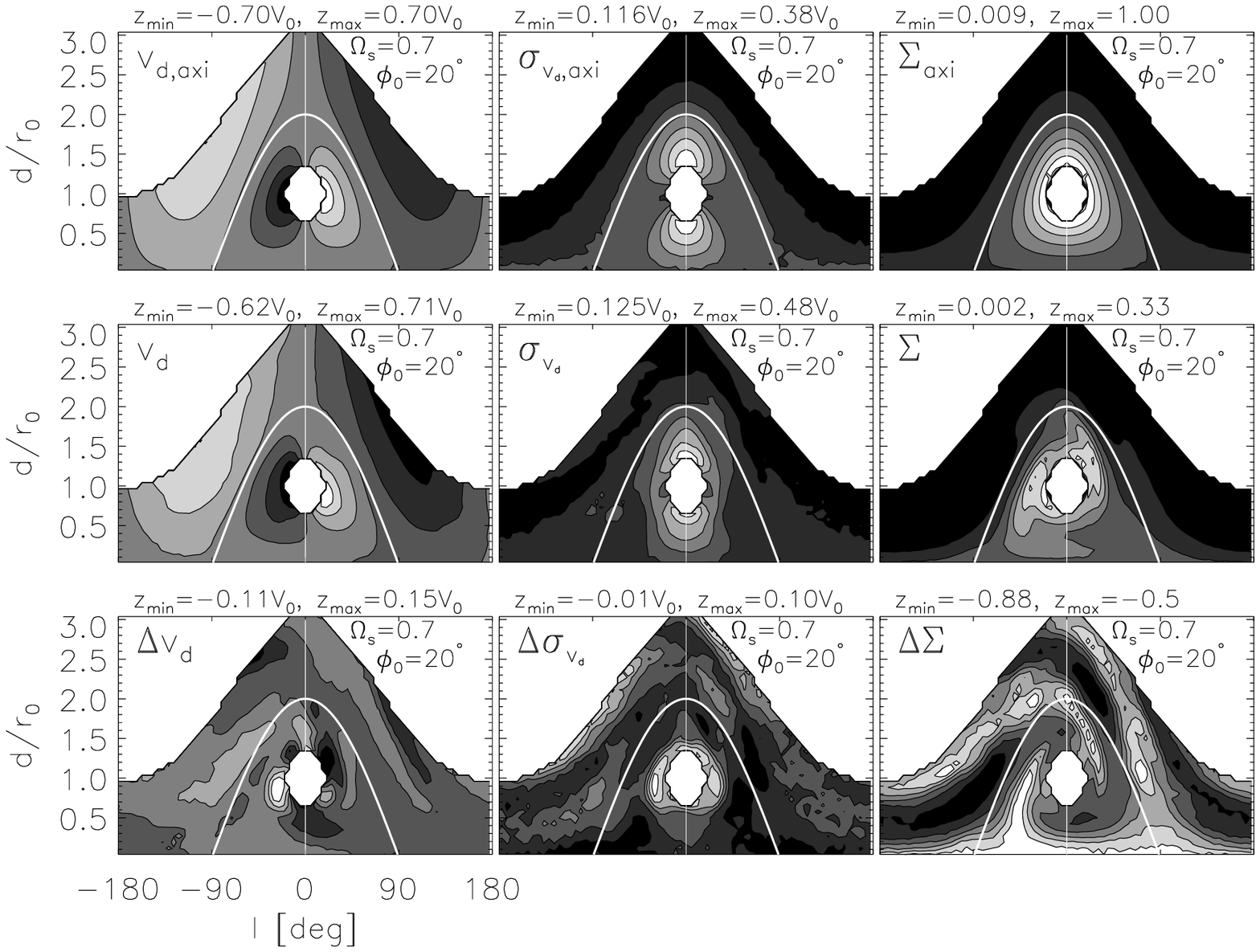}
\figcaption{
Pencil-beam maps of an axisymmetric disk (first row), same with an added 
two-armed spiral structure perturbation (second row), and the sole effect of
the spiral structure (third row). The pattern speed is $\Omega_s=0.7\Omega_0$ 
and the observer is at a solar phase angle of $\phi_0=20^\circ$. 
Columns from left to right show contour plots of 
the line-of-sight velocity $v_d$, its standard deviation $\sigma_{v_d}$, and 
the number density $\Sigma$. All of these are plotted versus the Galactic 
longitude $l$ (x-axis), and the heliocentric distance $d/r_0$ (y-axis). 
The white curve and vertical line are the projection of the solar circle and
the Galactic longitude $l=0^\circ$, respectively.
The number density plot in the x-y plane of this simulation 
was shown in figure \ref{fig:den_m2} (panel with $\Omega_s=0.7$). 
Clearly, the nonaxisymmetric structure caused by the effect of the spiral arms
is much better pronounced in the axisymmetric background subtracted values
(third row).
\label{fig:pbm20}
}
\end{figure*}

In general, changing the solar radius in a simulation with the same pattern speed 
is equivalent to 
changing the pattern speed and keeping the solar radius fixed. However, this is
exactly true only if the stellar density and velocity dispersion varied linearly with
radius. This is not the case in real galaxies; both of these are found to vary
exponentially with radius. \cite{lewis89} estimated the number density and radial
velocity dispersion scale lengths in the Milky Way to be $r_\rho=0.37r_0$, and 
$r_\sigma=0.9r_0$, respectively. Thus we need to perform different simulation runs 
when changing the pattern speed.

Note the disruption of the spirals near the 2:1 LRs (dashed circles). 
A rapid decrease of spiral strength at the 2:1 OLR was also observed by \cite{theis07}
in a galactic disk model consisting of solving numerically the Boltzmann moment 
equations. 
It has also been suggested by \cite{cont85} that strong (nonlinear) spiral structure 
cannot extend beyond the 4:1 ILR since at that location
the stellar orbits are not in phase with the imposed spiral. As pointed out by 
\cite{sellwood93}, however, this limited extent of the spirals found by 
\cite{cont85} is probably related to the restrictive assumptions they make in order to 
constructing self-consistent spiral structure. 
Another remarkable feature in the plots of figure \ref{fig:den_m2} is the 
overdensity of stellar orbits
just outside the 2:1 LRs (where they exist) and near the CR (dash-dotted circles).
In the case of the CR, the enhancement is due to the stable Lagrange points
$L_{14}$ associated with this resonance.

Assuming the primary spiral structure in the solar neighborhood is two-armed, the
question arises: How can we determine any spiral structure parameters, 
given our inconvenient position in the Galaxy?
One way to do this is by collecting a large number of stars with known velocities,
distances, etc, and constructing velocity and density maps by plotting these 
versus Galactic longitude, $l$ and heliocentric distance, $d$. 
In the next section we show what such maps would
look like when a two-armed spiral structure perturbs a stellar disk. We examine
different pattern speeds and solar positions with respect to a spiral arm. 
We call these maps "Pencil Beam Maps" or PBMs.

\section{Pencil Beam Maps (PBMs) of a Galactic disk perturbed by a two-armed 
spiral structure}
\label{sec:pbm}

To investigate the global structure of the Galaxy
we require accurate stellar velocities and distances. 
In a pencil-beam spectroscopic survey line-of-sight velocities can be measured
to great distances. On the other hand, proper motions are hard to measure for
stars farther than about two kpc from the Sun. Thus those cannot be used in our 
investigation.

For a complete kinematic study accurate distance estimates are also needed. 
Due to the large distances involved in a such survey trigonometric parallax 
measurements are not possible. Instead, photometric distances can be estimated
given accurate photometry. This way of computing distances, however, is 
hampered by the dust obscuration in the Galactic plane aside from several 
known windows, e.g., Baade's Window at $(l, b) = (0.9^\circ, -4^\circ)$. 
Another distance estimator is the use of standard candles such as Cepheids, etc.
Here we do not attempt to model the reddening resulting from dust extinction 
but present an idealized model as a first attempt to tackle this problem. A
future paper will be dedicated to a more detailed modeling. Due to this 
shortcoming our model can be directly applied only to the known low extinction
Galactic plane windows.
Like Baade s window, the Scutum window at $l = 27^\circ$ has low extinction and 
we can observe stars at ~10 kpc distances towards the inner disk. 
Clump giants of the intermediate-age and older population of the disk and 
thick disk will be abundant in these fields. This line-of-sight at $l = 27^\circ$ 
is tangent to the Scu-Cru spiral arm, with an AV extinction of about 3 mag at the 
distance of the spiral arm tangent point ($\sim6$ kpc). 
In the Scutum window, the HI and H$\alpha$ profiles clearly show the presence of
spiral arms \citep{madsen05}.

In figure \ref{fig:pbm20} we present PBMs of the line-of-sight velocity $v_d$ 
(left column), the corresponding velocity dispersion $\sigma_{v_d}$ (middle 
column), and the stellar number density $\Sigma$ (right column). 
This is a simulation of a galactic disk perturbed by a two-armed
spiral density wave moving with $\Omega_s=0.7\Omega_0$. To create the contour
plots in this figure we bin the disk in
Galactic longitude $l$ (x-axis), and heliocentric distance $d/r_0$ (y-axis) as 
seen from an observer at a solar orientation with respect to the concave 
spiral arm of $\phi_0=20^\circ$. This is in contrast to figure \ref{fig:den_m2}
where we present number density plots of a face-on view. The contours in the
first row in figure \ref{fig:pbm20} show the results of an axisymmetric disk, 
which is indicated by the subscript $``axi"$. In the second row the disk is 
perturbed by 
an $m=2$ spiral density wave. The third row in figure \ref{fig:pbm20} plots 
contours of the difference between the perturbed and axisymmetric disks for the
mean velocity and its dispersion: $\Delta v_d\equiv v_d-v_{d,axi}$ and 
$\Delta \sigma_{v_d}\equiv \sigma_{v_d}-\sigma_{v_d,axi}$. The number density,
on the other hand, is obtained as in figure \ref{fig:den_m2}: 
$\Delta \Sigma\equiv (\Sigma-\Sigma_{axi})/\Sigma_{axi}$
We showed a number density plot in the x-y plane of this particular simulation
in figure \ref{fig:den_m2} (panel with $\Omega_s=0.7$).  
In figure \ref{fig:pbm20}, however, we plot observables from a point of view
centered on the Sun, as pencil-beam surveys would see the Galaxy. 
The shaded contours are equally spaced
with darker color corresponding to lower density. On top of each panel we show
the minimum and maximum contour values (in units of $v_0$ in the case of the 
velocities and the velocity dispersion). As it is commonly accepted, the
Galactic longitude is zero in the direction of the Galactic center, with the 
Galactic anticenter at $l=\pm180^\circ$. The inner $0.3r_0$ disk has been removed 
everywhere as in the density plots in figure \ref{fig:den_m2}. 
The white curve in each panel represents the projection of the solar circle in a 
PBM and the vertical line shows the Galactic longitude $l=0^\circ$.  
  
\begin{figure*}
\epsscale{1.0}
\plotone{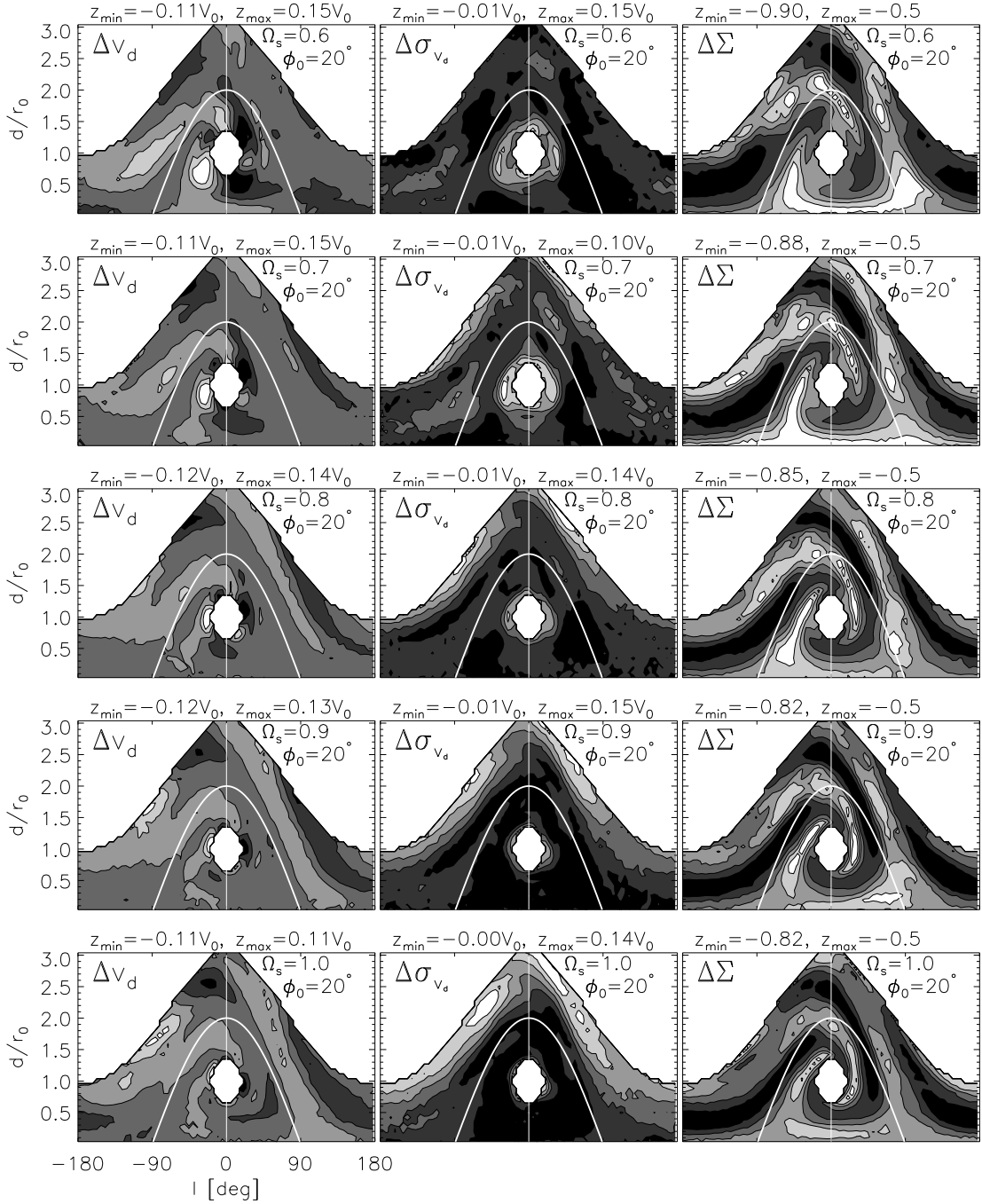}
\figcaption{
Pencil-beam maps showing variation of structure in the observables with the 
change in pattern speed. The solar angle is kept fixed at $\phi_0=20^\circ$ 
everywhere. In contrast to figure \ref{fig:pbm20} here all panels have the 
axisymmetric background
subtracted to reveal the symmetry of the spiral residuals. Note for the case of 
$\Delta\sigma_{v_d}$ the ring-like shape around the Galactic center for all pattern
speeds (middle column). These ''hot" rings are associated with the 2:1 ILR induced by
the spiral density wave. As the 2:1 OLR enters the disk (around $\Omega_s=0.9\Omega_0$,
see figure \ref{fig:den_m2}) strong features in $\Delta\sigma_{v_d}$ PBMs appear in
the outer disk. 
\label{fig:phi20}
}
\end{figure*}

Note that in the background subtracted PBMs (third row) the spiral structure 
is much better pronounced compared to the raw values (first row). 
Knowing the global Galactic potential is imperative to extracting information 
from a PBM. This presents a problem since the Milky Way potential is not very 
well known. We could, however, use our model axisymmetric data to subtract from the 
observational data. 

How well do we need to know the axisymmetric structure so that spiral features are
not wiped out? From the third row of figure \ref{fig:pbm20} we can estimate that
distance uncertainties of $\sim30\%$ do not prevent detection of spiral structure.
We also need line-of-sight velocity precision of $\sim20$ km/s and the 
axisymmetric potential must be known to within $\sim 10\%$.

What information about the spiral structure can we infer from PBMs such as
figure \ref{fig:pbm20}?
To answer this question we need to vary the parameters and look at how structure
in these PBMs changes. We do this in the following sections.

\subsection{Changing the spiral pattern speed}
\label{sec:om}

We would like to know how to infer the spiral pattern speed from structure in the 
PBMs. 
In figure \ref{fig:phi20} we plot the variation of PBMs with a change in the spiral
pattern speed in the range $0.6\leq\Omega_s/\Omega_0 \leq1.0$; the solar orientation
with respect to the concave spiral arm is kept fixed at $\phi_0=20^\circ$. 
This range places the Sun from just inside the 4:1 ILR to the CR.
In contrast to figure
\ref{fig:pbm20} here all panels have the axisymmetric background subtracted to 
reveal the symmetry of the spiral residuals. 
We are now looking for features that are strong enough to be detected in an actual 
pencil-beam survey. The $z_{min},z_{max}$ values indicated above each panel 
give the maximum error introduced by spiral structure in the otherwise axisymmetric
background disk. These values for the line-of-sight velocity $v_d$, and its 
standard deviation $\sigma_{v_d}$, are $\sim35$ km/s for $v_0=220$ km/s 
(left and middle columns in figure \ref{fig:phi20}), which is well above the 
resolution of upcoming radial velocity surveys ($<3$ km/s). 
Strong features showing marked variation with the change in pattern speed are 
apparent in all three observables: 
\newline
(1) In the case of $v_d$ high positive and negative velocity groups
resulting from the effect of the 2:1 ILR are found at 
$(l,d/r_0)\approx(-30^\circ,0.7)$ and $(l,d/r_0)\approx(25^\circ,1.3)$, respectively,
for $\Omega_s=0.6\Omega_0$ (top left panel). With the increase of pattern speed, 
these clumps spiral in a clockwise direction toward the Galactic center.  
\newline
(2) The standard deviation of the line-of-sight velocities, $\sigma_{v_d}$, which 
can also be described as the "heating" (or the random motions) of stars, 
peaks at a particular ring-like shape around the Galactic center for all pattern
speeds (middle column). These rings are associated with the 2:1 ILR induced by 
the spiral density wave. It is clear that the radii of these rings are changing with 
the change of the pattern speed and thus the location of the 2:1 ILR. 
Beyond the CR ($\Omega_s=\Omega_0$) this resonance falls inside the inner 
three kpc or inside the Galactic bulge. Thus, using the radius of this hot ring
to infer the location of the 2:1 ILR (and thus the pattern speed) is only valid 
if $\Omega_s$ values are in the range considered in figure \ref{fig:phi20}. 
\newline
(3) Lastly, the number density PBMs in figure \ref{fig:den_m2} (right column)
are also indicative of the changing pattern speed. The disruption of the spiral
arms near the 2:1 LRs and variation in spiral strength due to the encounter the 
second order resonances and CR, creates a large contrast in these axisymmetric 
background subtracted PBMs. 
Many of these features can be used in addition to the information extracted from
the velocities. 

All of the features described above can be used to identify the location of the
2:1 ILR and thus the patter speed. As we mentioned, however, if the solar circle 
is placed at or beyond the CR, the 2:1 ILR falls inside the Galactic bulge.
Consequently, the features created by it disappear. Fortunately, just as this
happens the 2:1 OLR enters the Galactic disk (our disks extend to a radius of 
$2r_0$) and similarly to the 2:1 ILR case, resonant features are created, this 
time in the outer parts of the disk. 

\begin{figure*}
\epsscale{1.0}
\plotone{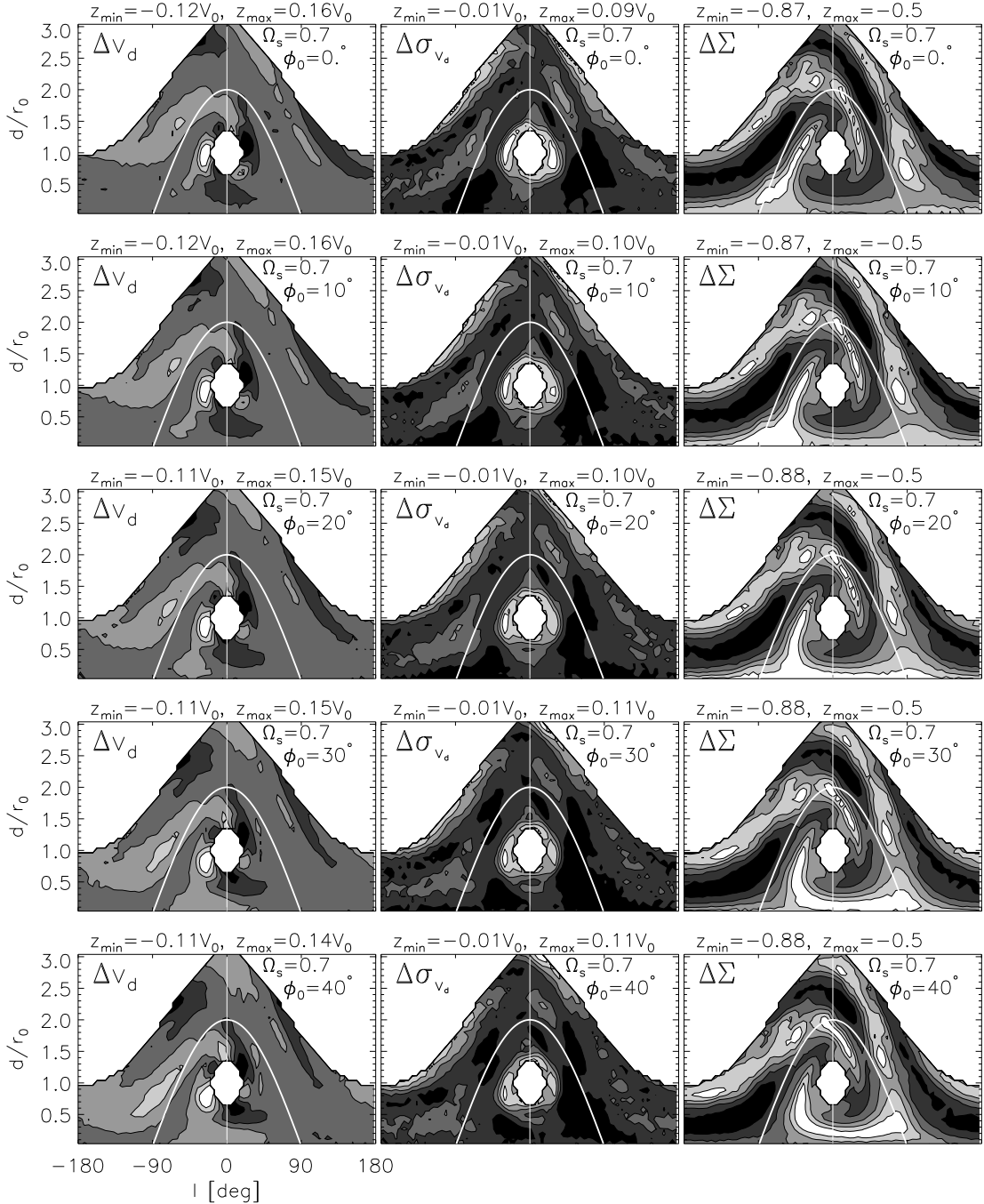}
\figcaption{
Pencil-beam maps showing variation of structure in the observables with solar
orientation at a fixed pattern speed $\Omega_s=0.7\Omega_0$. As in figure 
\ref{fig:phi20} all panels have the axisymmetric background subtracted. 
The solar phase angle is varied in the range $\phi_0=[0^\circ,40^\circ]$. 
The negative velocity stream found in the $\Delta v_d$ PBMs (left column) at 
$(l,d/r_0)\approx(0^\circ,0.5)$ moves toward positive longitude with the change
of phase. This is a strong feature ($\sim-20$ km/s) and can be used to estimate 
$\phi_0$. Even more indicative of changes in the solar phase angle are the 
number density PBMs (right column). For example pencil-beam observations at 
$\pm45^\circ$ would be drastically different depending on the phase angle.
\label{fig:om0.7}
}
\end{figure*}

\subsection{Changing the Sun's orientation with respect to a spiral arm}

How can we infer the Sun's azimuth with respect to the galactocentric line 
passing through the intersection of the Solar circle and a concave spiral arm?
To find out we plot PBMs of simulation runs with the same pattern
speed and different solar phase angle. Figure \ref{fig:om0.7} shows such plots
for a fixed $\Omega_s=0.7\Omega_0$ and a solar phase angle changing from top to 
bottom in the range $\phi_0=[0^\circ,40^\circ]$. 

Similarly to figure \ref{fig:phi20} we now look for strong features in the
three observables that can be used to estimate $\phi_0$:
\newline
(1) Inspection of the left column of figure \ref{fig:om0.7} reveals a negative 
velocity stream which changes position with a change of phase angle. 
For $\phi_0=0$ (top left) this feature is centered on 
$l\approx0^\circ$ at a heliocentric distance of $d/r_0\approx0.5$. As the angle is
increased this steam moves to larger longitudes roughly preserving its distance 
from the Sun. Note that the high positive and negative features discussed in 
the context of $v_d$ in figure \ref{fig:phi20}, do not vary as the angle is 
changed since the pattern speed is kept fixed.
\newline
(2) The line-of-sight velocity dispersion (middle column of figure \ref{fig:phi20})
does not seem to be particularly useful for constraining the solar phase angle.
\newline
(3) Finally, the structure in the number density PBMs shows prominent variation with 
the change in solar angle. For example pencil-beam observations at $\pm45^\circ$ 
would be drastically different depending on the phase angle. 

In an actual survey we would first try to infer the position of the inner or outer
LR as discussed in Section \ref{sec:om} and thus find the pattern speed.

\section{Four-armed spiral structure}
\label{sec:m4}

\begin{figure*}
\epsscale{1.0}
\plotone{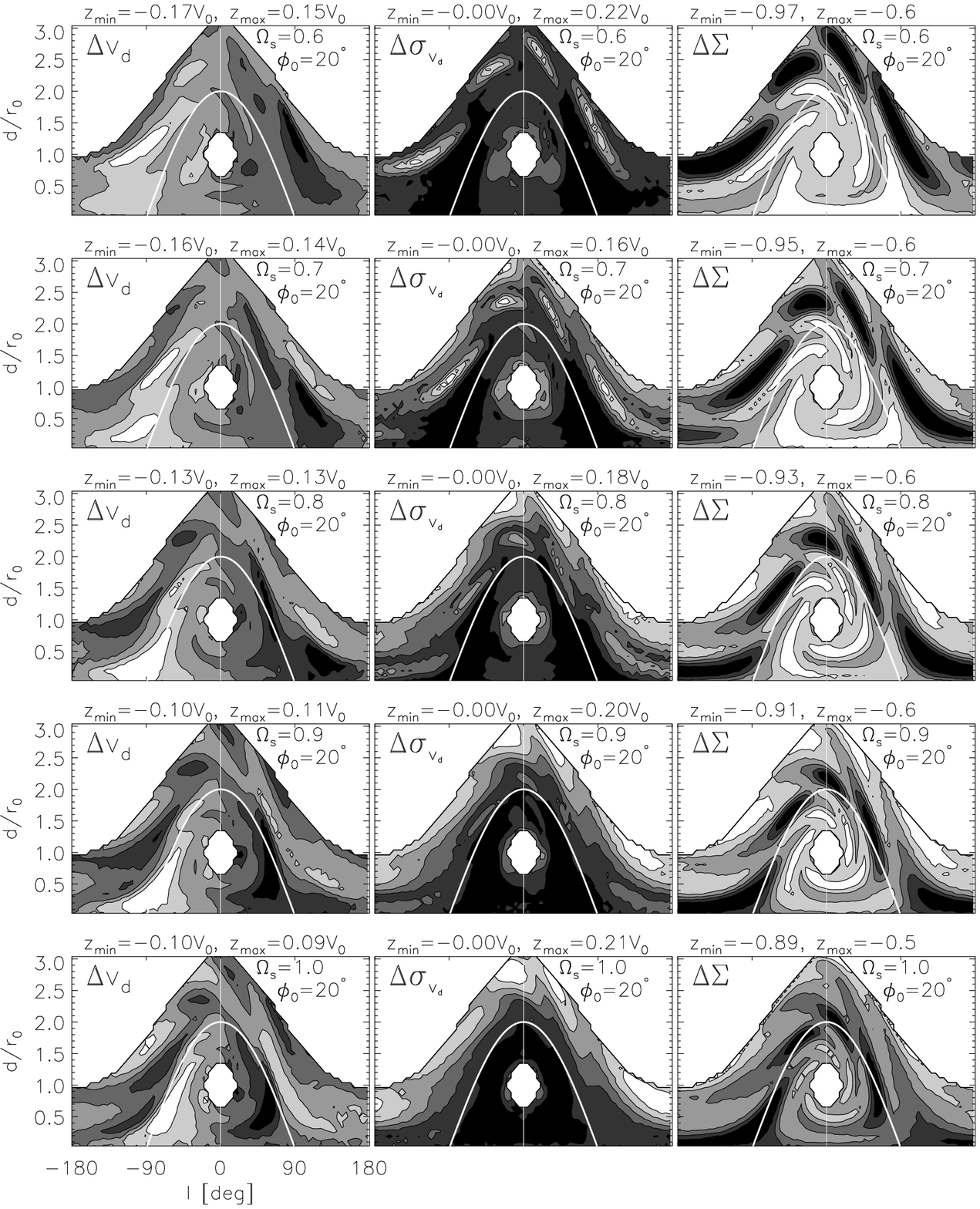}
\figcaption{
Same as figure \ref{fig:phi20} but in the case of a four-armed spiral structure.
Pattern speed changed from top to bottom in the range $\Omega_s=[0.6,1.0]\Omega_0$
and the solar orientation with respect to a concave arm is kept fixed at 
$\phi_0=20^\circ$. Inspecting $\Delta v_d$ and $\Delta\Sigma$ (left and right columns) 
it is clear that a pencil beam radial velocity observation along 
Galactic longitude of $l=-90^\circ$ can unambiguously distinguish between $m=2$ 
and $m=4$ structure.
\label{fig:phi20m4}
}
\end{figure*}

So far we have only discussed simulations involving a two-armed spiral density wave
perturbation. In this section we show the effect of a four-armed structure, make 
comparison with the two-armed case and suggest a way to distinguish between the two.

Figure \ref{fig:phi20m4} shows PBMs of a galactic disk, similarly to figure 
\ref{fig:phi20}, but perturbed by a four-armed spiral structure. 
As in figure \ref{fig:phi20} pattern speed changed from top to bottom in the range 
$\Omega_s=[0.6,1.0]\Omega_0$ and the solar orientation with respect to a concave
arm is kept fixed at $\phi_0=20^\circ$. In this case the first order resonances are 
the 4:1 ILR/OLR which occur at $\Omega_s=0.65\Omega_0,1.35\Omega_0$, respectively.

As expected, more structure is apparent in all PBMs in the case of the four-armed 
structure. Note that the hot rings in $\Delta\sigma_{v_d}$ present in the case of the 
two-armed structure (figure \ref{fig:phi20}) are also apparent in the four-armed case
although not as pronounced. The reason for this is the fact that the 
2:1 ILR which causes these is a second order when $m=4$. These hot rings can be 
used in both the two- and four-armed cases to estimate $\Omega_s$ but are expected to 
be much stronger for $m=2$. 
Inspecting $\Delta v_d$ and $\Delta\Sigma$ (left and right columns) in figure 
\ref{fig:phi20} and \ref{fig:phi20m4} it is clear that pencil-beam observations
along Galactic longitudes $l=0^\circ$ or $l=-90^\circ$, for example, can 
unambiguously distinguish between $m=2$ and $m=4$ spiral structure, as the 
oscillation frequency doubles when $m=4$. 

\section{Conclusion}

Upcoming Galactic disk surveys will reveal the age, composition and phase 
space distribution of stars within the various Galactic components. These stellar 
excavations will provide essential clues for understanding the structure, 
formation and evolution of our Galaxy. 
To facilitate the interpretation of the huge amounts of data resulting from these
surveys, Galactic disk models, such as the one presented here, are needed to 
interpret the observations.

We have investigated how the Milky Was spiral structure 
parameters, such as pattern speed and solar phase angle, can be estimated in a
deep all-sky survey. We performed a series of test-particle simulations of a 
warm galactic disk approximating the disk kinematics of the Milky Way.
We considered both two- and four-armed spiral structure and suggested a way to 
distinguish between the two using velocity and number density maps. 

We found that the axisymmetric potential needs to be known to $\sim10\%$, 
line-of-sight velocities to $\sim20$ km/s, and distance uncertainties need to be
less than $\sim30\%$.
The mean line-of-sight velocity and the velocity dispersion are affected by up to 
$\sim35$ km/s
which is well within the detectable limit for forthcoming radial velocity surveys.
Pattern speed can be constrained by a hot ring at the 2:1 ILR in both two- and
four-armed spiral structure. To distinguish between the two, however, we also 
need information related to the velocities and stellar number density.
If the pattern speed is such that the 2:1 ILR is hidden inside the Galactic bulge
the 2:1 OLR would be present in the outer Galaxy and thus can equivalently be
used to estimate the pattern speed. Once the pattern speed is known the solar angle
can be estimated from the number density variation with heliocentric distance;
$\phi_0$ is also reflected in the $v_d$ PBMs.

Future work needs to address the issue of how to obtain the axisymmetric 
background potential needed to subtract from the observational data as discussed
at the end of Section \ref{sec:pbm}. Also, it is important to know what type of 
tracer stars are needed that would allow the estimation of photometric parallaxes 
with errors less than $\sim30\%$, and the distribution of those stars.

While here we only considered steady state spiral stricture, other theories
of spiral structure, such as transient and swing-amplified spirals, need to be
investigated as well.
It has also been suggested that the Galaxy harbors two sets of spiral structure 
moving at the same \citep{lepine01} or different \citep{naoz07,mq06} pattern 
speeds. We expect in all those cases it will be again resonant features to
relate to the pattern speed and solar angle. In the case of non-steady state
spirals, however, the structure in the PBMs will vary with integration time
and interpretation will become more complicated.
We refer these cases to a future study.

It is also known that the Milky Way is a barred galaxy. The simulations performed
here do not include the influence of the bar. This is not necessarily a 
shortcoming since most of the features in the PBMs we use to infer spiral structure
parameters are caused by resonances and, unless a resonance overlap with the
central bar exists in the same location, those would not be different when a 
bar is included in the simulations. Future work should also look at this problem.

\acknowledgements
Support for this work was in part provided by National Science Foundation 
grant ASST-0406823, and the National Aeronautics and Space Administration
under Grant No.~NNG04GM12G issued through the Origins of Solar Systems Program.


\begin{thebibliography}{}

\bibitem[\protect\astroncite{{Bissantz} et~al.}{2003}]{bissantz03}
{Bissantz}, N., {Englmaier}, P., and {Gerhard}, O.: 2003,
\newblock {\em \mnras} {\bf 340}, 949

\bibitem[\protect\astroncite{{Blitz} and {Spergel}}{1991}]{blitz91}
{Blitz}, L. and {Spergel}, D.~N.: 1991,
\newblock {\em \apj} {\bf 379}, 631

\bibitem[\protect\astroncite{{Contopoulos}}{1985}]{cont85}
{Contopoulos}, G.: 1985,
\newblock {\em Comments on Astrophysics} {\bf 11}, 1

\bibitem[\protect\astroncite{{De Simone} et~al.}{2004}]{desimone04}
{De Simone}, R., {Wu}, X., and {Tremaine}, S.: 2004,
\newblock {\em \mnras} {\bf 350}, 627

\bibitem[\protect\astroncite{{Dehnen}}{1999}]{dehnen99}
{Dehnen}, W.: 1999,
\newblock {\em \apjl} {\bf 524}, L35

\bibitem[\protect\astroncite{{Dehnen}}{2000}]{dehnen00}
{Dehnen}, W.: 2000,
\newblock {\em \aj} {\bf 119}, 800

\bibitem[\protect\astroncite{{Dias} and {L{\'e}pine}}{2005}]{dias05}
{Dias}, W.~S. and {L{\'e}pine}, J.~R.~D.: 2005,
\newblock {\em \apj} {\bf 629}, 825

\bibitem[\protect\astroncite{{Drimmel} and {Spergel}}{2001}]{drimmel01}
{Drimmel}, R. and {Spergel}, D.~N.: 2001,
\newblock {\em \apj} {\bf 556}, 181

\bibitem[\protect\astroncite{{Fux}}{2001}]{fux01}
{Fux}, R.: 2001,
\newblock {\em \aap} {\bf 373}, 511

\bibitem[\protect\astroncite{{L{\'e}pine} et~al.}{2001}]{lepine01}
{L{\'e}pine}, J.~R.~D., {Mishurov}, Y.~N., and {Dedikov}, S.~Y.: 2001,
\newblock {\em \apj} {\bf 546}, 234

\bibitem[\protect\astroncite{{Lewis} and {Freeman}}{1989}]{lewis89}
{Lewis}, J.~R. and {Freeman}, K.~C.: 1989,
\newblock {\em \aj} {\bf 97}, 139

\bibitem[\protect\astroncite{{Lin} et~al.}{1969}]{lin69}
{Lin}, C.~C., {Yuan}, C., and {Shu}, F.~H.: 1969,
\newblock {\em \apj} {\bf 155}, 721

\bibitem[\protect\astroncite{{Madsen} and {Reynolds}}{2005}]{madsen05}
{Madsen}, G.~J. and {Reynolds}, R.~J.: 2005,
\newblock {\em \apj} {\bf 630}, 925

\bibitem[\protect\astroncite{{Martos} et~al.}{2004}]{martos04}
{Martos}, M., {Hernandez}, X., {Y{\'a}{\~n}ez}, M., {Moreno}, E., and
  {Pichardo}, B.: 2004,
\newblock {\em \mnras} {\bf 350}, L47

\bibitem[\protect\astroncite{{Minchev} et~al.}{2007}]{mq07b}
{Minchev}, I., {Nordhaus}, J., and {Quillen}, A.~C.: 2007,
\newblock {\em \apjl} {\bf 664}, L31

\bibitem[\protect\astroncite{{Minchev} and {Quillen}}{2006}]{mq06}
{Minchev}, I. and {Quillen}, A.~C.: 2006,
\newblock {\em \mnras} {\bf 368}, 623

\bibitem[\protect\astroncite{{Minchev} and {Quillen}}{2007}]{mq07a}
{Minchev}, I. and {Quillen}, A.~C.: 2007,
\newblock {\em \mnras} {\bf 377}, 1163

\bibitem[\protect\astroncite{{Naoz} and {Shaviv}}{2007}]{naoz07}
{Naoz}, S. and {Shaviv}, N.~J.: 2007,
\newblock {\em New Astronomy} {\bf 12}, 410

\bibitem[\protect\astroncite{{Paczynski} et~al.}{1994}]{paczynski94}
{Paczynski}, B., {Stanek}, K.~Z., {Udalski}, A., {Szymanski}, M., {Kaluzny},
  J., {Kubiak}, M., and {Mateo}, M.: 1994,
\newblock {\em \aj} {\bf 107}, 2060

\bibitem[\protect\astroncite{{Quillen} and {Minchev}}{2005}]{qm05}
{Quillen}, A.~C. and {Minchev}, I.: 2005,
\newblock {\em \aj} {\bf 130}, 576

\bibitem[\protect\astroncite{{Sellwood}}{1993}]{sellwood93}
{Sellwood}, J.~A.: 1993,
\newblock {\em \pasp} {\bf 105}, 648

\bibitem[\protect\astroncite{{Shaviv}}{2003}]{shaviv03}
{Shaviv}, N.~J.: 2003,
\newblock {\em New Astronomy} {\bf 8}, 39

\bibitem[\protect\astroncite{{Vorobyov} and {Theis}}{2007}]{theis07}
{Vorobyov}, E.~I. and {Theis}, C.: 2007,
\newblock {\em ArXiv e-prints} 709

\bibitem[\protect\astroncite{{Weinberg}}{1992}]{weinberg92}
{Weinberg}, M.~D.: 1992,
\newblock {\em \apj} {\bf 384}, 81

\end{thebibliography}
\end{document}